\documentclass{article}
\usepackage{icrctc07}

\title{Muons and Neutrinos 2007}
\shorttitle{Muons \& Neutrinos}

\authors{Thomas Gaisser} 
\shortauthors{T.K. Gaisser}
\afiliations{Bartol Research Institute, Department of Physics and Astronomy,
 University of Delaware, Newark, DE 19716, U.S.A.}
\email{gaisser@bartol.udel.edu}

\abstract{This paper is the written version of the rapporteur talk on 
Section HE-2, {\em muons and neutrinos}, presented at the 30th International Cosmic Ray
Conference, Merid\'{a}, Yucatan, July 11, 2007.  Topics include atmospheric muons
and neutrinos, solar neutrinos and astrophysical neutrinos as well as calculations
and instrumentation related to these topics.
}
\begin{document}
\maketitle

\section{Introduction}
There were 5 sections of contributed papers on muons and neutrinos with a total of
107 papers distributed as shown in Table~\ref{table1}.  The most active category
is HE 2.3, {\em astrophysical neutrinos}.  Particularly in this area, there were also many 
papers presented in OG 2.5 sessions, on {\em high-energy astrophysical neutrinos}.
I include discussion of these topics to the extent necessary to present a coherent
overview of the field as of mid-year 2007.

\begin{table}[htb]
\label{table1}
\caption{Papers on muons and neutrinos}
\begin{tabular}{l|l|l}\\ \hline
Session & Topic & \# \\ \hline
HE 2.1 & Muon experiments & 17 \\
HE 2.2 & Observations of solar & \\ 
 & \& atmospheric $\nu$ & 16 \\
HE 2.3 & Observations of & \\
& astrophysical $\nu$ & 37 \\
HE 2.4 & Theory and simulations & 19 \\
HE 2.5 & New experiments & \\
&\& instrumentation & 18 \\ \hline
\end{tabular}
\end{table}

\begin{figure}
\vspace{-.5cm}
\begin{center}
\noindent
\includegraphics[width=0.5\textwidth]{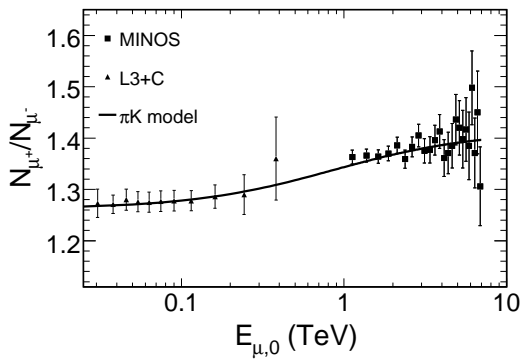}
\end{center}
\caption{Muon charge-ratio from Ref.~\protect\cite{MINOS}.  The line
``$\pi K$" model corresponds to a fit to the ratio of $\mu^+/\mu^-$
from Eq.~\ref{muons}.
}
\label{fig1}
\end{figure}

\section{Atmospheric muons}
Muons are the gold standard of cosmic-ray physics because they are well-measured
and their physical origin in the atmospheric cosmic-ray cascade is well-understood.
A summary of measurements is included in this conference in Ref.~\cite{Kempa}.
Muons are penetrating and relatively abundant in all terrestrial particle detectors.  They
are therefore a potential source of background, and at the same time they are
useful for detector calibration.  One use of cosmic-ray muons is as a survey tool,
sometimes called muon tomography.  A classic example is the survey of the Second Pyramid of Giza
and the search for hidden chambers~\cite{Alvarez}.  The status of a similar 
investigation of the Pyramid of the Sun at Teotihuacan was presented at this conference~\cite{ArturoM-R}.
The detector is integrated and ready for installation during 2008 in the tunnel
that goes under the pyramid.

An important new result presented at this conference is the measurement of the muon charge ratio in the 
far detector of MINOS~\cite{MINOS,Mufson,Goodman}.  Muons that reach the detector at its depth of 
2070 meters-water-equivalent (m.w.e.) have energies at the surface in the range of 1-7 TeV,
depending on zenith angle.  Fig.~\ref{fig1} from Ref.~\cite{MINOS} shows muon charge ratio
increasing in the energy range from $E_\mu < 100$~GeV to $E_\mu > 1$~TeV.  The potential significance
of this result can be understood from the energy-dependence of Eq.~\ref{muons}.

An approximate, first-order expression for the muon intensity 
for $E_\mu>100$~GeV is~\cite{TKG}
\begin{eqnarray}
\label{muons}
\phi_{\mu^\pm}&=&{\phi_0(E_\mu)\over 1\;-\;Z_{NN}}\times \\ \nonumber
&&\left\{ 
{A_{\pi\mu}Z_{N\pi^\pm}\over 1 + B_{\pi\mu}E_\mu\cos\theta/\epsilon_\pi}\right. \\ \nonumber
 &&+\;\left.
{A_{K\mu}Z_{NK^\pm}\over 1+B_{K\mu}E_\mu\cos\theta/\epsilon_K}
\right\}, \nonumber
\end{eqnarray}
where $Z_{N\pi}$ and $Z_{NK}$ are spectrum weighted moments for production
of pions and kaons, $\phi(E)\equiv E\;{\rm d}N/{\rm d}E$ and $\phi_0(E_\mu)$
is the intensity of primary cosmic-ray nucleons evaluated at the energy of the muon.
The kinematic factors are $A_{\pi\mu}\approx 0.67$, $A_{K\mu}\approx 0.25$ (including
the branching ratio for $K\rightarrow \mu\nu$), $B_{\pi\mu}\approx 1.07$ and
$B_{K\mu}\approx 1.13$.

Zatsepin and Kuz'min pointed out long ago~\cite{ZK} the potential of a measurement of
the atmospheric muon flux as a function of zenith angle for measuring the relative
importance of kaon to pion production in hadronic interactions.  The angular dependence
arises from the denominators of the two terms in Eq.~\ref{muons} and the numerical
values of the critical energy parameters, $\epsilon_\pi\approx 115$~GeV and
$\epsilon_K\approx 850$~GeV.  In the TeV range, the kaon contribution is relatively
more important near the vertical than at large angle.  Thus the angular dependence
of the muon flux is sensitive to the ratio $Z_{NK}/Z_{N\pi}$.  The angular dependence
of the muon charge ratio provides information on the relative importance of kaons and
pions separated by charge.  
By fitting the data of Fig.~\ref{fig1} to the charge ratio calculated from Eq.~\ref{muons},
the MINOS group find
\begin{equation}
{Z_{N\pi^+}\over Z_{N\pi^+}+Z_{N\pi^-}}\;=\;0.55
\end{equation}
and
\begin{equation}
{Z_{NK^+}\over Z_{NK^+}+Z_{NK^-}}\;=\;0.67.
\label{ratio}
\end{equation}
The ratio $(Z_{NK^+}+Z_{NK^-})\;/\;(Z_{N\pi^+}+Z_{N\pi^-})$ has been kept fixed
at its standard value.  The fit uses the data of Refs~\cite{MINOS,L3}, which
have data binned both in $E_\mu$ and $\cos\theta$.  The preliminary
measurement of the charge ratio in the shallow MINOS near detector~\cite{deJong}
(not shown) is consistent with the L3+C data.
The large value of the $K^+/K^-$ ratio reflects the importance of forward associated
production ($p\rightarrow \Lambda\; K^+$) which is amplified in the spectrum-weighted
moment by the steep primary cosmic-ray spectrum because the $K^+$ carries
on average a significant fraction of the beam energy.

Two other papers from MINOS are noteworthy as examples of the use of cosmic-ray
muons for calibrating deep detectors.  More than 20 million
muons were measured (after cuts) in the MINOS far detector over a three-year
period from August 2003 to August 2006.  One 
analysis uses the shadow of the moon to determine
the angular resolution and absolute pointing of the far detector~\cite{Grashorn2}.
Both the resolution and the absolute pointing are $0.3^\circ\pm 0.05^\circ$.
The moon's shadow is seen  at the level of $4\;\sigma$.  The shadow of the sun
is also seen, but at a somewhat
lower significance, in part because of bending of the parent cosmic
rays in the solar magnetic field.

Another paper~\cite{Grashorn1} presents the analysis of seasonal 
variations of
the underground muon rate observed in the MINOS far detector.  The observed
rate of muons is correlated with temperature by
\begin{equation}
{\Delta R_\mu\over <R_\mu>}\;=\;\alpha^T\times{\Delta T_{\rm eff}\over <T_{\rm eff}>},
\label{seasonal}
\end{equation}
where $T_{\rm eff}$ is an average of the temperature weighted by 
the probability of meson production, which peaks at altitudes around
15~km for trajectories near the vertical.  
This correlation is
explained in the classic paper of Barrett et al.~\cite{classic} 
as a consequence expansion of the atmosphere when temperature increases.
The correlation is large
for muons with $E_\mu >> \epsilon_\pi$ where there is a competition
between decay and re-interaction of the parent pion.  At lower energies,
most pions decay before interacting for any temperature.  The observed 
rate, $R_\mu$, shows a seasonal variation of $\approx\pm 2$\% and corresponds
to a value of $\alpha_T = 0.87\pm 0.03$, consistent with earlier measurements
by MACRO~\cite{MACROseasonal} and AMANDA~\cite{AMANDAseasonal}.

\section{Atmospheric neutrinos}
Similar equations to Eq.~\ref{muons} 
describe the flux of atmospheric neutrinos at high energy.
However, the kinematic factors differ in an important way.  Because its mass
is close to that of the pion, the muon carries most
of the energy in the $\pi\rightarrow\mu\nu$ decay.  The decay $K\rightarrow \mu\nu$
is more nearly symmetric.  As a result, while the kinematic parameters for
$K\rightarrow\nu$ are nearly equal to those for $K\rightarrow\mu$, they
are quite different for pions.  In particular, $A_{\pi\nu}\approx 0.088$
as compared to $A_{\pi\mu}\approx 0.67$.  As a consequence, the dominant
contribution to neutrinos with $E_\nu>100$~GeV is from kaons, and the
effect of the large charge ratio for kaons has a strong influence on the
atmospheric neutrino spectrum at high energy.  The tendency is to harden
the TeV neutrino spectrum and to increase the ratio $\nu_\mu / \bar{\nu}_\mu$.

It is now widely accepted that the deficit of atmospheric muon-neutrinos
with its pathlength and energy dependence is the result of neutrino oscillations.
At this conference, there were only two papers concerning calculation of the
flux of atmospheric neutrinos.
Reference~\cite{GDBarr} estimates and tracks the various sources of
uncertainty through the calculation in order to evaluate the systematic
uncertainty in the flux of atmospheric neutrinos as a function of neutrino
energy.  The contribution of Honda et al.~\cite{Hondaetal}, focuses on
evaluation of relatively small effects such as variation with solar
cycle and the effect mountainous overburden above the detector.  Ref.~\cite{Hondaetal}
is based on a revised calculation~\cite{HondaPub} that uses a new model of meson
production in hadronic interactions~\cite{SanukiPub} based on comparison to
measurements of atmospheric muons.  The new Honda et al. neutrino flux is now
closer to the Bartol neutrino flux~\cite{GDBarr,Barretal} at high energy than
the earlier calculation~\cite{Honda2004}.
In both models now, the ratio $Z_{pK^+}/Z_{pK^-}$ is large, consistent
with the interpretation of the increase in the $\mu^+/\mu^-$ ratio in the
TeV region discussed above.  The corresponding effect here is that the ratio
$\nu/\bar{\nu}$ is large ($\sim 1.7$) in the TeV range).  Moreover, the 
TeV neutrino flux is relatively high because of the importance of forward associated
production on the steep cosmic-ray spectrum.  (An independent confirmation of
the relatively high atmospheric neutrino flux in the TeV range comes from
the work of Ref.~\cite{Concha}, which assumes the best fit oscillation parameters
and unfolds the atmospheric neutrino spectrum from the Super-K measurements.)
The implications for atmospheric neutrinos of the MINOS measurement of the muon charge
ratio have yet to be investigated in detail, however.  It should be possible to use the MINOS
data to
reduce the uncertainty in existing calculations of the flux of atmospheric neutrinos
in the TeV region and above.

\section{Neutrino oscillations}

Super-Kamiokande has been restored to its full complement of over 11,000
50 cm photomultipliers plus an outer veto detector and has been operating 
since July 12, 2006 as Super-Kamiokande-III.  Reference~\cite{Miura}
reviews the history of Super-K, which began operation in April 1996 and
announced the discovery of oscillations of atmospheric neutrinos in 1998~\cite{SK1998}.
The phase of operation up to the accident in November, 2001 is SK-I.  The detector was
repaired and operated with some 5000 PMTs redistributed to provide uniform but 
sparser coverage.  SK-II ran for three years, starting October, 2002.
Preliminary results of SK-III are in agreement
with SK-I and SK-II. 

A series of Super-K papers at this conference presented
preliminary results of the combined analysis of SK-I (1489 days) and SK-II (804 days).
Atmospheric neutrino results, for example, were presented~\cite{Takenaga} in the
same format as the main SK-I paper~\cite{SK2005}.  The plots of zenith angle show
a ratio of $(\nu_e+\bar{\nu}_e) / (\nu_\mu+\bar{\nu}_\mu)$ that is significantly 
higher than expectation for sub-GeV neutrinos from all directions, and a deficit of
multi-GeV neutrinos from below ($\sim$10,000 km) but consistent with expectation
from above ($\sim$15 km).  The angular distribution of the electron neutrinos
has the expected (no oscillation) shape.  The results are fully consistent with
two-flavor $\nu_\mu\leftrightarrow\nu_\tau$ oscillations with transition
probability
\begin{equation}
P_{\nu_\mu\leftrightarrow\nu_\tau}=
\sin^2(2\theta_{23})\times\sin^2\left [
1.27{\delta m^2({\rm eV}^2)L_{\rm km}\over E_{\rm GeV}}\right].
\label{osc}
\end{equation}
The dip at the first oscillation minimum in $L/E$ is seen~\cite{dip} for 
atmospheric
neutrinos~\cite{Higuchi}. and there is no
evidence yet for three-flavor effects such as non-zero $\theta_{13}$. 

The MINOS group also presented their results for neutrino oscillations using
the NuMI muon-neutrino beam from Fermilab~\cite{Sousa}.  The results, already
published,~\cite{MINOS-PRL} show a deficit of muon neutrinos in the far detector
relative to the near detector over a distance of $735$~km that is consistent
with the results of the Super-K atmospheric neutrino result.  The MINOS far detector
can also measure $\nu_\mu$-induced upward muons with charge separation.
Although statistics are limited, they see a deficit of lower-energy 
neutrino-induced muons consistent with the Super-K oscillation 
parameters~\cite{Mufson711}.
An interesting feature of neutrino-induced muons in a magnetized detector
is that the charge ratio can be measured.  The charge ratio is opposite to
that for atmospheric muons because positive mesons ($\pi^+$ and $K^+$) decay to
$\mu^+$ and $\nu_\mu$ (which produce $\mu^-$), 
while negative mesons give $\bar{\nu}_\mu$ (which produce $\mu^+$).

During the time that Super-K II operated with half the density of PMTs 
as compared to Super-K I, 
new reconstruction algorithms were developed that allowed sensitivity similar to
that of the original detector.  Now that the detector has been restored to its
full complement of PMTs, the better algorithms make it possible to lower the
energy threshold~\cite{Smy}.  Super-K III currently is operating with 100 per cent 
trigger efficiency down to $5$~MeV, which is in the transition region from 
matter dominated to vacuum oscillations for solar neutrinos.

\section{Astrophysical neutrinos (low energy)}
Neutrinos from SN1987A in the Large Magellanic Cloud are so far the only neutrinos 
detected~\cite{Kam,IMB} from outside the
solar system.  A network of several deep detectors continues to monitor the sky for bursts
of neutrinos from nearby stellar collapses.  New upper limits
on the rate of stellar collapse in the Milky Way Galaxy based on non-observation of neutrino
bursts are summarized in Table~\ref{tab2}.

\begin{table}[htb]
\label{tab2}
\caption{Limits on supernova rates in the Milky Way Galaxy (events per year at 90\% c.l.).
The Super-K limit includes SMC and LMC.}
\begin{tabular}{l|l|l}\\ \hline
Experiment & Exposure & Limit \\ \hline
S-K~\cite{Ikeda} & 2589 d & 0.30 \\
LVD~\cite{LVD} & 4919 d & 0.17 \\
Baksan~\cite{Baksan} &  22 yr & 0.10 \\ \hline
\end{tabular}
\end{table}

It is also possible to search for a diffuse flux of relic neutrinos from past supernova explosions.
The spectrum of these relic neutrinos peaks at a few MeV and falls quickly with increasing
energy~\cite{Totani}.  The process
\begin{equation}
\bar{\nu}_e\;+p\;\rightarrow\;n\;+\;e^+
\label{inverseBetaDecay}
\end{equation}
with its relatively large cross section is the preferred channel for this search~\cite{Kaplinghat}.
The convolution of the cross section, which increases with energy, and the spectrum of relic neutrinos
may be above the expected background from atmospheric $\bar{n}$ in a window of energy
from $\sim$10 to 20 MeV~\cite{Beacom}.  Current limits from Super-K~\cite{Malek} are close to the
signals expected from various models, as shown in Ref.~\cite{Iida} at this conference.
The possibility~\cite{Beacom} of adding Gadolinium to 
tag recoil nucleons from the process of Eq.~\ref{inverseBetaDecay} was mentioned at this conference
in Ref.~\cite{Smy}.  A test of this method with a 2.4 liter container of GdCl$_3$ and
a radioactive source is described in Ref.~\cite{Watanabe}.

\subsection{Astrophysical neutrinos (high energy)}
The most promising channel to use in the search for astrophysical neutrinos of high
energy ($>$~TeV) is neutrino-induced muons
because the effective volume of the detector is amplified by the muon range.
The average energy loss rate of a muon per $X$(g/cm$^2$) of material traversed is
\begin{equation}
{{\rm d}E\over {\rm d}X}\;=\;-a\;-bE,
\label{Eloss}
\end{equation}
where $\epsilon = a/b\sim 0.5$~TeV is the characteristic energy above which
stochastic losses (bremsstrahlung and nuclear interactions) begin
to dominate the energy loss.  
The corresponding average muon range is
\begin{equation}
X\;\approx\;{1\over b}\times ln\left({E_{\mu,0} + \epsilon
\over E_{\mu,{\rm min}} + \epsilon}\right),
\label{Range}
\end{equation}
where $E_{\mu,0}$ is the muon energy at production and
$E_{\mu,{\rm min}}$ is the threshold energy of the detector for muons.
The muon range can be several kilometers or more in water or ice.

When only the muon is detected, however, there is only an average relation between the
visible energy of the muon and that of the neutrino that produced it.
The fraction of the neutrino energy carried by the
muon varies from event to event, and the track is only partially contained.  Moreover,
for $E_\mu>>\epsilon$ there are large fluctuations 
in the amount of visible energy deposited as the high-energy muon passes through
the detector~\cite{Misaki}.  Nevertheless, because the relation between energy deposition
of the muon as it passes through the detector and its total energy
is well understood (as well as the relation between the energy of
the muon and that of the neutrino that produced it), it is straightforward to 
derive the parent neutrino spectrum from a measurement of neutrino-induced
muons given sufficient statistics.  
From a Monte Carlo simulation of the
detector response one has to identify a set of measurable quantities
that depend on energy deposition in the detector.  An unfolding procedure
can then be used to reconstruct the parent spectrum.  The error analysis
must account for the large fluctuations from event to event.  
A prescription using the photon density along the muon track as the
energy-dependent observable for reconstruction and unfolding
the atmospheric neutrino spectrum in IceCube is given in Ref.~\cite{Dima}.
This is a natural choice given the physics of muon energy loss
described by Eq.~\ref{Eloss}.

\begin{figure}[t]
    \includegraphics[width=0.5\textwidth]{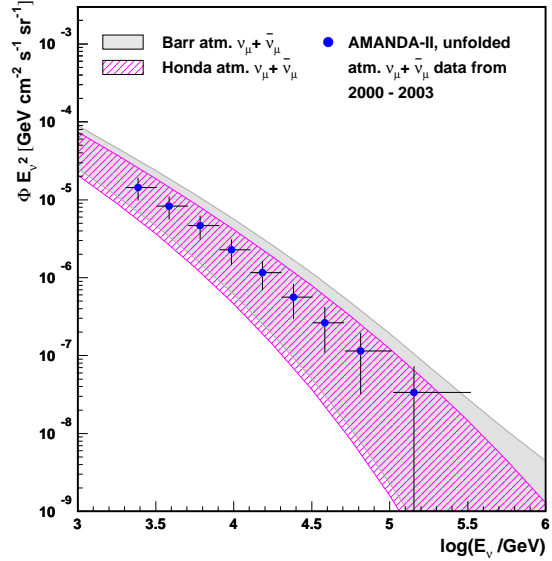}
    \caption{Unfolded spectrum of atmospheric neutrinos by AMANDA-II~\cite{Kirsten}
compared to calculations of Refs.~\cite{Barretal,Honda2004}.
}
    \label{unfolded}
\end{figure}

The high-energy tail of the spectrum of atmospheric neutrinos constitutes the
background for searches for neutrinos from astrophysical sources.  Atmospheric
neutrinos also serve as the calibration beam.  Atmospheric neutrinos are sufficiently
well-understood in the multi-TeV range so that successful reconstruction of their
spectrum can be considered as a prerequisite to any search for astrophysical neutrinos.
As an example, Fig.~\ref{unfolded} shows the atmospheric neutrino spectrum
derived by an unfolding procedure from AMANDA data taken from  2000-2003~\cite{Kirsten}.

\begin{figure*}
\begin{center}
\includegraphics [width=0.9\textwidth]{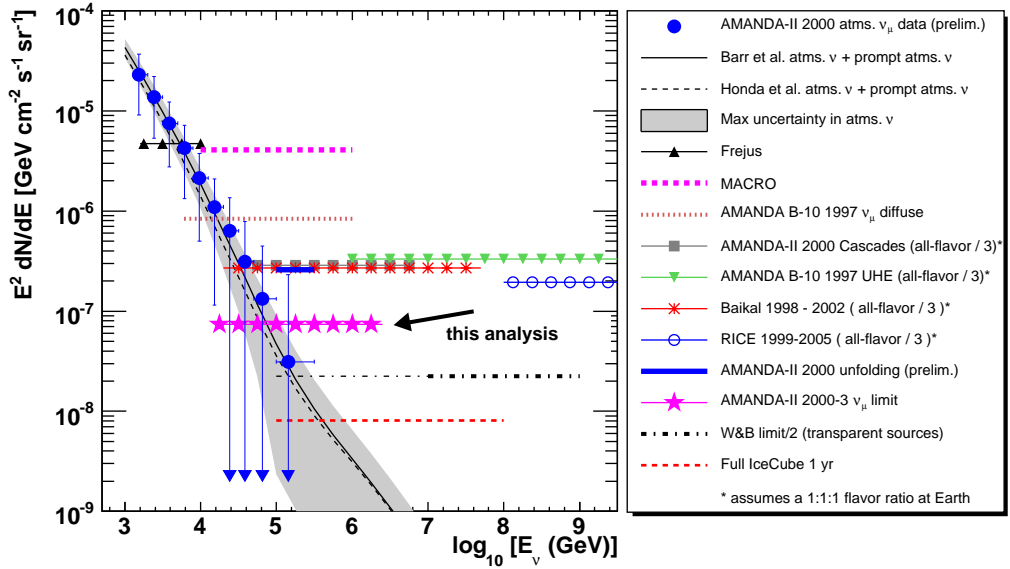}
\end{center}
\caption{Limit on the diffuse flux of astrophysical 
neutrinos from AMANDA-II~\cite{Jessica}.}\label{Jessica}
\end{figure*}

An important remaining uncertainty in the atmospheric neutrino flux at high energy
is the level of the contribution from prompt neutrinos.  These are neutrinos
from the decay of charmed hadrons which have a harder spectrum 
than neutrinos from decay of pions and kaons, which are suppressed at
high energy because the parent mesons tend to interact rather than decay.
The prompt contribution to the atmospheric lepton flux can be represented by
adding a third term to the right hand side of Eq.~\ref{muons} of the form
$$
{A_{C}Z_{N,C}\over 1+B_{C}E\cos\theta/\epsilon_C},
$$
where the subscript "{\it C}" represents a charmed hadron,
E is the lepton energy ($\mu$, e, $\nu_\mu$ or $\nu_e$) and
$\epsilon_C \sim 2 - 9\times 10^7$ for a range of charmed hadrons
with significant leptonic branching ratios~\cite{Kampert}.
For muon neutrinos with $E_\nu > \epsilon_K = 850$~GeV the spectrum
gradually steepens from $E^{-2.7}$ to $E^{-3.7}$ while the
spectrum of prompt neutrinos continues to reflect the primary cosmic-ray spectrum
until $E_\nu =\epsilon_C\sim 3\times 10^7$~GeV.  The crossover energy
depends on the amount of charm production in hadronic interactions
at high energy ($Z_{N,C})$, which is highly uncertain, particularly in the
fragmentation region.  For a model with a significant contribution
of intrinsic charm~\cite{Bugaev} the neutrinos from charm decay become
the dominant component of atmospheric $\nu_\mu$ for $E_\nu > 100$~TeV~\cite{TG-Honda}.
For the much steeper spectrum of $\nu_e$ the crossover of the charm component is
around 3~TeV.

{\bf Diffuse search.}   High-energy astrophysical neutrinos are expected to be 
produced by interaction 
of high-energy accelerated particles with gas or electromagnetic radiation in or near the
sources.  Generally the particle beams, and hence the produced neutrinos, are expected
to have a harder energy spectrum than the background atmospheric neutrinos.
A standard benchmark for high-energy neutrinos of extra-galactic origin is 
the Waxman-Bahcall limit~\cite{WaxBah}, which assumes an $E^{-2}$ differential spectrum.
The normalization of the Waxman-Bahcall limit for $\nu_\mu +\bar{\nu}_\mu$
at Earth after accounting for oscillations is 
$E^2{\rm d}N_\nu /{\rm d}E < 2.2\times 10^{-8}$ GeV\,cm$^{-2}$s$^{-1}$sr$^{-1}$.
To exceed this level would require the existence of cosmic accelerators opaque to
the particles they accelerate.  The limit might also be relaxed to some extent
at lower energy in the case of steeper source spectra.

Figure~\ref{Jessica} shows current limits on a diffuse flux of neutrinos
with an $E^{-2}$differential energy spectrum from AMANDA.  The figure
illustrates several points.  The search labeled "this analysis" 
looks for neutrino-induced muons generated in the ice and rock 
below the detector.  This is the energy region and the signature
for which existing large detectors in deep water or ice are optimized.
Such a search is limited to neutrinos
with $E_\nu<1$~PeV because the Earth absorbs neutrinos with higher energy.

In the PeV energy region and above a diffuse signal would be
dominated by events near the horizon, where the target length
is maximized without absorbing the neutrinos.  Here one is
generally looking for events characterized by large and concentrated depositions
of energy, either radiating $\nu_\mu$-induced muons or
cascades from interactions of $\nu_e$ or $\nu_\tau$
in or near the detector.  Fig.~\ref{Jessica}
includes the limit from the Baikal experiment in this energy range.
The "all-flavor" limit is divided by three on the plot
to make it comparable with the limits on $\nu_\mu$ alone.
A preliminary value of $2.4\times 10^{-7}$ GeV\,cm$^{-2}$s$^{-1}$sr$^{-1}$
was presented at this conference for the upper limit of all neutrino flavors 
from AMANDA in the "UHE" energy range $10^5 < E_\nu < 10^9$~GeV~\cite{Lisa}.  This is
approximately a factor of three below the Baikal limit and only slightly
higher than the limit from upward muons~\cite{Jessica} in the energy
range around $1$~PeV where both limits apply.

The "diffuse" limit on any particular model spectrum is obtain by
convolving the spectrum with the detector response.  In the
case of the diffuse limits for an $E^{-2}$ spectrum, the plotted
limit is a horizontal line on a plot of $E^2{\rm d}N/{\rm d}E$
extending over the energy region that gives 90\% of the signal.
In general, a separate limit must be calculated for each assumed
model spectrum.  Ref.~\cite{Lisa} gives a useful
table of models and sensitivities that specifies which
models are inconsistent at 90\% confidence level with
the present AMANDA UHE diffuse limit.  Several early models
of neutrino production in AGN are ruled out,
including, for example Ref.~\cite{MPR}, while others (e.g. \cite{Stecker,Mannheim})
are still viable.

In the case of searches for UHE neutrinos with optical detectors the signal 
would be characterized by a large amount of light in the detector.  Since
the events will be from above and from the sides, an important background
is from bundles of muons generated by high-energy cosmic rays cascades in
the atmosphere.  Showing that this physical background is well-understood
(for example by comparing simulations with data at various cut levels) is
needed to demonstrate understanding of the detector response.  A feature 
used to discriminate between signal and background in Ref.~\cite{Lisa}
is the number of optical modules with multiple hits versus single hits.
Muon bundles from cosmic-ray cascades tend to produce less light per particle, and the source of
the light is somewhat diffuse as compared to the intense and concentrated
burst of light from a single particle with energy in the PeV range or higher.
The signal would produce more multiple hits.

\begin{table*}
\begin{center}
\begin{tabular}{lccccccc} \hline
& & & &$E_\nu>1$ TeV & & $E_\nu>5$ TeV & \\
Source name & Dia($^\circ$) & Vis & $\epsilon_\nu$ (TeV) & $N_{\rm src}$ & $N_{\rm atm}$ & 
$N_{\rm src}$ & $N_{\rm atm}$ \\ \hline
A: RX J0852.0-4622 & 2.0 & 0.83 & 1.19 & 11 & 104 & 4.2 & 21 \\
A: RX J1713.7-3946 & 1.3 & 0.74 & 1.25 & 11 & 41 & 4.6 & 8.2 \\
B: LS 5039 (INFC) & 0.1 & 0.57 & 1.01 & 0.5 & 2.5 & 0.2 & 0.5 \\
C: HESS J1303-631 & 0.3 & 1.0 & 0.21 & 1.6 & 11 & 0.3 & 2.1 \\
D: Vela X & 0.8 & 0.81 & 0.84 & 16 & 23 & 10 & 4.6 \\
D: HESSJ1825-137 & 0.5 & 0.57 & 4.24 & 8 & 9.3 & 3.7 & 1.8 \\
D: Crab Nebula & $<$0.1 & 0.39 & 1.72 & 5.8 & 5.2 & 1.9 & 1.1 \\ \hline
\end{tabular}
\caption{TeV galactic $\gamma$-ray sources from the H.E.S.S. catalog~\cite{HESS} with
corresponding neutrino rates calculated for 5 years operation of KM3NeT (1 km$^3$ instrumented
volume)~\cite{Kappes}.
}\label{galactic}
\end{center}
\end{table*}

{\bf Point sources}.  Particularly luminous and/or nearby sources of neutrinos
should eventually emerge above the diffuse atmospheric background.  Likely candidates
are the subset of gamma-ray sources in which the gamma-rays are hadronic in
origin, from decay of neutral pions produced in interactions of accelerated protons and nuclei
in or near the sources.  The kinematic relation between $\pi^0\rightarrow\gamma\gamma$
and $\pi^+\rightarrow\mu^+\nu_\mu$ provides a close connection between neutrinos
and gamma-rays if the photons are not significantly absorbed in the sources.
Examples of potential sources are Active Galactic Nuclei (AGNs), Gamma-ray Bursts (GRBs)
and supernova remnants and active compact objects in our galaxy.  Since it is still 
not known which gamma-ray sources are hadronic, identification neutrinos
from sources of gamma rays (and/or electromagnetic radiation in other wavelengths)
is the central goal of high energy neutrino astronomy. 

As an example, it is interesting to consider likely sources of high-energy neutrinos in
our local galaxy.  In the Northern hemisphere, visible from Antarctica, the Cygnus
region is of particular interest~\cite{Abdo}.  A systematic survey of the
sensitivity of a future kilometer-cube neutrino detector in the Mediterranean
to potential galactic sources visible from the North was given in Ref.~\cite{Stegmann}.
 (A more
detailed discussion of the analysis is given in Ref.~\cite{Kappes}.)  
Table~\ref{galactic}
summarizes results for those H.E.S.S. sources with spectra for which a break 
energy (where the spectrum steepens) has been determined.  The corresponding
neutrino spectrum would steepen at a somewhat lower energy than the observed
steepening of the gamma-ray spectrum and
about a factor of 40 lower than the energy at which the parent proton spectrum
steepens.  Assumptions of the calculation are that the observed gamma-ray
spectrum is entirely hadronic in origin, produced by interaction of an
accelerated spectrum of protons with gas in or near the source and that there
is no absorption of gamma-rays in the source.  A ratio at production of
$\nu_e:\nu_\mu:\nu_\tau\;=\;1:2:0$ is assumed with a flavor ratio at Earth of
$1:1:1$.  The neutrino effective area of a km$^3$ detector in the Mediterranean
for the $\nu_\mu+\bar{\nu}_\mu$ channel
is calculated in some detail to obtain the expected number of events for
source ($N_{\rm src}$) and background ($N_{\rm atm}$) by convolution with
the spectrum of the source and with the atmospheric neutrino spectrum.  
However, efficiencies for 
event selection and reconstruction are not accounted for.
 
The results summarized in Table~\ref{galactic} nicely illustrate 
some important features of point
source searches with kilometer-scale neutrino telescopes.  The second
and third columns of
the table give the diameter of each gamma-ray source and the
fraction of the time it is below the horizon.  Typically, the
source sizes exceed the resolution of H.E.S.S. and are comparable to or
larger than the resolution of the neutrino telescope.  If the characteristic
neutrino break energy is $\ge 1$~TeV or higher, the signal to background
ratio improves at higher energy, so the ability to measure
a signal related to energy will be important.  For neutrino sources that steepen
in the TeV region, the signal/background improves by about a factor of two
if the threshold can be raised from 1 to 5 TeV.

Expected rates are low, and signal/background is less than or comparable
to unity depending on the size of the source.  Techniques such
as ``source stacking" will therefore be important~\cite{Becker}
to improve the significance.  Similar conclusions about signal/background
for such galactic source can be inferred from Ref.~\cite{Carr}.
Hadronic models are disfavored for several of the types of sources listed.
Prime candidates are A: shell-type SNRs and C: TeV $\gamma$-ray sources with
no counterparts at other wavelengths.  Pulsar wind nebulae (D) and binary systems (B)
are more often explained with electromagnetic models, although hadronic
models exist.

\begin{figure}[t]
    \includegraphics[width=0.5\textwidth]{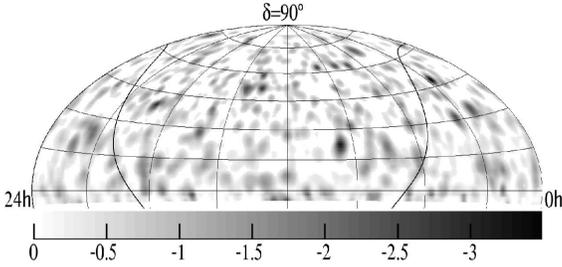}
    \caption{Preliminary sky map from Ref. \cite{Braun} showing 
log$_{10}$(p) for an unbinned point source search with AMANDA-II
in 2005.}
    \label{IC9}
\end{figure}

Traditional searches use a bin size around the source optimized for the source
size and point spread function of the detector.  In Ref.~\cite{Stegmann,Kappes}, for example,
the bin size is $1.6\times\sqrt{\sigma^2_{\rm PSF}+\sigma^2_{\rm src}}$.  
Unbinned likelihood procedures that improve sensitivity by using energy-dependence
and time clustering (as well as direction) are discussed for ANTARES in Ref.~\cite{Aguilar},
for AMANDA in Ref.~\cite{Braun} and for IceCube in Ref.~\cite{Finley}.  Ref.~\cite{Braun}
finds an improvement in sensitivity and discovery potential for AMANDA-II
of 30\% compared to the binned search.  The ANTARES analysis~\cite{Aguilar}
finds a greater improvement, up to a factor of two or more in some cases.  Figure~\ref{IC9}
illustrates the kind of confidence level map that results from an unbinned point source search.

{\bf Variable sources}.  For sources known to be variable in electromagnetic
radiation, e.g. in X- or $\gamma$-radiation, the significance of a small number
of neutrinos from the direction of that source could be greater
if they occur at the same time as flares.  Assessing the extra significance
depends on the extent to which the pattern of flaring is understood.
Ref.~\cite{Bernardini} proposes a method in which directions to sources such as specific AGN
known to be variable in electromagnetic radiation are searched for time clusters of
neutrinos on various time scales.  When a significant fluctuation above
background is found, a check is made to see if the source is in a high state
in electromagnetic radiation at the same time.  If not (or if the EM data are not
available) a time-clustering algorithm is compared with a large set of Monte Carlo
data samples from the selected set of sources to look for an excess over background.
A similarly motivated approach~\cite{Porrata} looks for correlations on various time scales
using known properties of the atmospheric neutrino background as part of the analysis.

The most straightforward approach to making use of variability in electromagnetic
radiation from a potential neutrino source is to look for a correlation in
the historical record between, for example, flares from blazars and times
of neutrinos from the direction of the same sources.
For coincidence with a gamma-ray telescope with nearly
continuous coverage of a large part of the sky (e.g. Milagro or Tibet) this is a good
approach~\cite{JGoodman}.  
For telescopes with a limited field of view, however 
(such as VERITAS, MAGIC, H.E.S.S.), the telescope will most likely be looking
elsewhere when a neutrino signal occurs.  If, as is likely, the neutrino events
are not significantly above atmospheric background on their own, then no signal can be claimed.
One way to address this asymmetry is to send an alert when a pre-specified
condition is satisfied by the neutrino detector, which is continuously sensitive
to the hemisphere below the detector.  The gamma-ray telescope 
can then slew to the selected source and see if it is flaring.  
A test of such a neutrino-triggered
"Target of Opportunity" (ToO) arrangement for a pre-selected set
of sources was reported at this conference for
AMANDA and MAGIC~\cite{ToO}.  

Another possibility is to define an alert as a group neutrinos from the same direction
within a pre-selected time window for any direction in the sky.  Such a possibility
is described in Ref.~\cite{Marek} where it is proposed to send an alert to
optical cameras that can quickly point to the direction defined by the
group of neutrinos.  In this way it might be possible to discover the onset of
an optical supernova or a GRB afterglow, which could elevate the significance
of the neutrino observation from a chance coincidence of several atmospheric
neutrinos to an identified astrophysical neutrino event.

\begin{figure}[t]
    \includegraphics[width=0.5\textwidth]{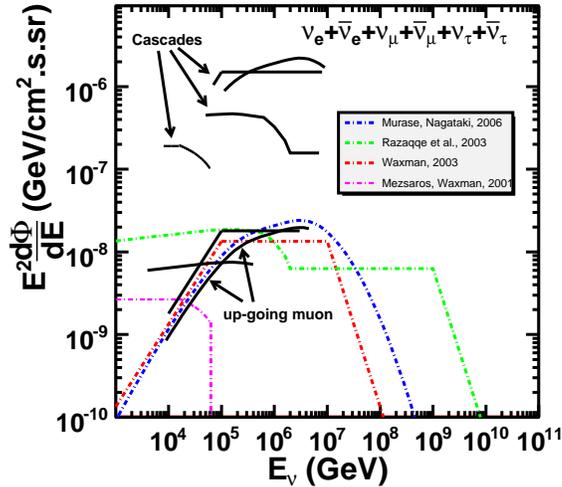}
\caption{AMANDA limits on neutrinos from GRBs~\cite{Marek}.  See text
for discussion.}
\label{Taboada}
\end{figure}

{\bf Gamma-ray bursts}.   
Neutrinos associated with gamma-ray bursts would have
both a time tag and a location which would make the detection of even a small number
of such neutrinos significant.  Limits on neutrinos associated with 
gamma-ray 
bursts using AMANDA has been published recently~\cite{Kyler,Ignacio} and presented at this
conference in Ref.~\cite{Marek}.  The most sensitive search was in the
$\nu_\mu$-induced muon channel~\cite{Kyler}, which used 400 hundred bursts
reported by BATSE and IPN3 between 1997 and 2003.  
No neutrinos were observed.  Fig.~\ref{Taboada} shows a comparison
of limits with models.  The three lines labeled ``up-going muon"
are limits for the models with the corresponding shapes.
Thus the model of Ref.~\cite{Razz} is ruled out and the model
of Ref.~\cite{Murase} is marginally incompatible with the limit.
A 3$\sigma$ upper limit is set at 1.3 times the level predicted in
the model of Waxman and Bahcall~\cite{WB-GRB}.  With IceCube the
sensitivity for detection of neutrinos from GRBs will rapidly improve.
IceCube is operating now with 22 strings and is expected to have
36 to 40 strings in operation by the time GLAST turns on in 2008.
Assuming GLAST will observe some 200 GRBs per year over the whole
sky, it is estimated~\cite{Marek} that observation of 70 bursts in
the Northern hemisphere without associated neutrinos would be in 
conflict with the model of Ref.~\cite{WB-GRB} at the $\sigma$ level.
This level of sensitivity should be possible with an exposure equivalent
to one year of full IceCube.

\begin{figure}[t]
    \includegraphics[width=0.5\textwidth]{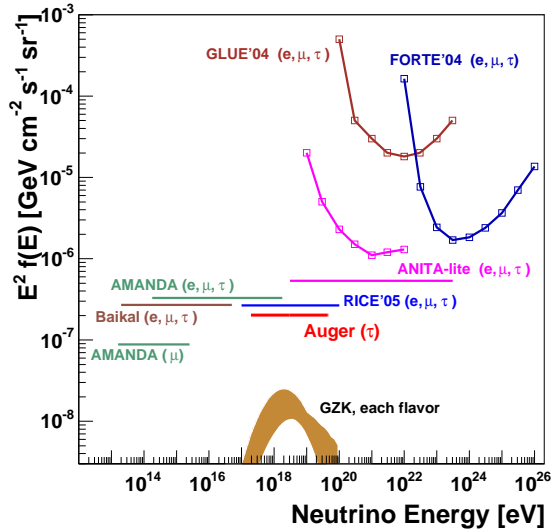}
    \caption{Figure from Ref. \cite{Blanch} showing upper limits
from various experiments 
assuming $\nu_e:\nu_\mu:\nu_\tau = 1:1:1$ at the detector.
See text for a discussion of this figure.
}
    \label{GZK}
\end{figure}

\section{Cosmogenic neutrinos}  
There is now a growing consensus that the primary cosmic-ray
spectrum becomes steeper above $5\times 10^{19}$~eV~\cite{Cronin,Watson,Teshima}.  
This is generally
attributed to the "GZK" effect~\cite{Greisen,ZatKuz} of energy loss
as particles interact with photons of the microwave background
radiation during propagation from sources at cosmological distances.
A lack of high energy particles could, however, also be due to a lack
of sources capable of accelerating particles to energies of
 $\sim 10^{20}$~eV.  In any case, the number and spectrum of cosmogenic
neutrinos is a key to the origin of the highest energy cosmic rays.
Current measurements have not yet reached the sensitivity to
detect cosmogenic neutrinos at the expected levels, as shown
in Fig.~\ref{GZK} from Ref.~\cite{Blanch}.

Limits shown in Fig.~\ref{GZK} come from several types
of detectors, which are sensitive to different ranges of
energy and to different combinations
of neutrino flavors.  The Auger limit~\cite{Blanch} is for the $\nu_\tau$
channel for a period from January 2004 to December 2006
that corresponds to one year of operation of the 
full detector.  The limit is shown for the
energy region that would generate 90\% of the signal
for an $E_\nu^{-2}$ differential spectrum, which
overlaps well with the expected spectrum of
cosmogenic neutrinos.  A similar limit
from Hi-Res~\cite{HiRes} is essentially at the same level
as the Auger limit.  The air-shower limits are based on
searches for atypical horizontal showers as discussed in the next section. 

For uniformity, limits from other experiments
are shown in Fig.~\ref{GZK} assuming an equal mixture of the three neutrino flavors
at Earth.  Limits from the optical 
detectors, AMANDA~\cite{AMANDA1,Jessica} and 
Baikal~\cite{Baikal2}, are at lower energy.  Limits from
radio detectors include RICE~\cite{RICE} in the ice at the South Pole,
ANITA-Lite~\cite{AnitaL}, a balloon-borne radio detector
looking for neutrinos interacting in the Antarctic
ice sheet, and FORTE~\cite{FORTE}, searching for radio pulses
from neutrino interactions in the Greenland ice mass.  
GLUE~\cite{GLUE} looks for
microwave signals of neutrino interactions in the Moon.

There are several calculations of the spectrum of cosmogenic neutrinos,
which vary depending on assumptions about the spectrum and cosmological
evolution of the cosmic-ray sources.  The band shown in Fig.~\ref{GZK} is
a range based on calculations by two groups~\cite{Engeletal,Allardetal}.
The calculation of Ref.~\cite{Allardetal} was discussed at this conference
in~\cite{Busca} for two different models of the primary cosmic-ray composition.
The result in the high-energy peak region ($E_\nu\sim 10^{18}$~eV) is rather
independent of the composition.

\begin{table*}[thb]
\begin{center}
\begin{tabular}{lcccl} \hline
Detector & Number of OMs & enclosed vol. (m$^3$) & depth (m.w.e.) & status \\ \hline
Baikal (NT200+) & 230 & $2\times 10^6$ & 1100-1310 & operating \\
AMANDA & 677 & $1.5\times 10^7$ & 1350-1850 & operating \\
ANTARES & 900 & $1\times 10^7$ & 2050-2400 & 2007/2008 \\
IceCube & 1320 & $1.8\times 10^8$ & 1350-2250 & 2007 \\
 & 4800 & $10^9$ & 1350-2250 & 2011 \\
KM3Net & $\sim$10,000 & km$^3$ & 2300-3300 (NEMO) & design study \\
&	&km$^3$& 3000-4000 (NESTOR) & \\
& & km$^3$ & 1400-2400 (ANTARES site) & \\ \hline
\end{tabular}
\caption{Parameters of existing and proposed neutrino neutrino telescopes in water and
ice.}
\label{table4}
\end{center}
\end{table*}

\section{Neutrino detectors and techniques}

I conclude with a summary of the status of large neutrino detectors
which have the primary aim of finding high energy ($\ge$TeV) astrophysical
neutrinos and identifying their sources.  The discussion is organized
by detection method.  I do not include here the densely instrumented
detectors such as MINOS and Super-K, which are aimed primarily at
study of neutrino oscillations and (in the case of Super-K) low energy
neutrinos and proton decay.

Because of oscillations, neutrinos from astrophysical sources are expected
to consist of comparable numbers of all three neutrino flavors after propagation
from distant sources.  At production (whether in the atmosphere or in an
astrophysical source) the production of $\nu_\tau$ is strongly suppressed
relative to $\nu_\mu$ and $\nu_e$.  For this reason, identification of
$\tau$-neutrinos would be a signal of astrophysical neutrinos.  The signature
of a tau neutrino interaction is expected to become recognizable at
high energy where the produced $\tau$-lepton has a measurable decay
length.  The decay length is
\begin{equation}
\Gamma c \tau_\tau\;=\;49\;{\rm m}\times E_\nu({\rm PeV}).
\label{tau}
\end{equation}
Several papers at this conference focus on the phenomenology of
$\tau$-neutrinos in the context of new experiments and the search
for cosmogenic neutrinos.

\begin{figure}[t]
    \includegraphics[width=0.5\textwidth]{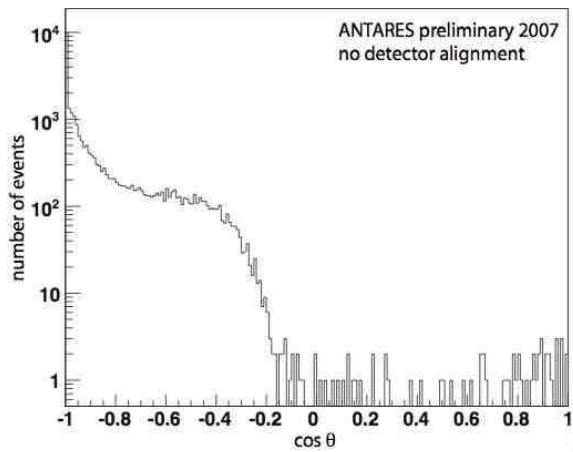}
    \caption{Figure from Ref. \cite{Kouchner} showing angular
distribution of reconstructed muons in 5 lines of ANTARES.
}
    \label{AntaresAngular}
\end{figure}

{\bf Optical detectors}
Antares and IceCube build on the optical techniques of Baikal and AMANDA.
Generally the water detectors have less scattering and therefore
superior ability at track reconstruction and angular resolution than
ice, while being subject to higher background noise rates due to
radioactivity and bioluminescence.  Fig.~\ref{AntaresAngular} from Ref.~\cite{Kouchner}
shows the preliminary angular distribution of reconstructed muon tracks.
The isotropic distribution of muons induced by atmospheric neutrinos
emerges from the background of downward muons already slightly above the
horizon.  This achievement reflects a combination of long scattering length
and depth of the detector.  At a depth of 2475 m, the intensity of
penetrating atmospheric muons is almost an order of magnitude lower
than at the top of IceCube (1450 m ice).
At the time of the conference (July 2007), ANTARES had 5 lines operating in the ice.
All 12 lines with a total 900 of optical modules are now in place and full operation of
the detector is set to begin early in 2008.  The status of various
neutrino telescopes that use the optical Cherenkov technique is summarized
in Table~\ref{table4}.

IceCube is currently operating with 22 strings and 1320 digital optical modules (DOMs)
at depths of 1450 - 2450 m in the ice at the South Pole~\cite{Karle}.  Results reported
at this conference were from data taken with the 9 string configuration that operated
during 2006.  First observations of atmospheric neutrinos with IceCube-9 have been
published~\cite{Pretz}.  The plan is to complete IceCube deployment over the next four
austral summer seasons to its full size.  The detector will operate and accumulate data
while deployment is being completed.  IceCube includes a surface component, IceTop~\cite{IceTop},
which ran with 16 stations during 2006 and is currently operating with 26 stations during 2007.
Each station consists of two ice Cherenkov tanks, each of which is viewed by two DOMs.
IceCube including IceTop constitutes a 3-dimensional air shower array that can study
primary cosmic rays, including primary composition, from PeV to EeV energy.  AMANDA is
now integrated into IceCube as a densely instrumented sub-array~\cite{Andreas}
There is a common event builder
so that every reconstructed event in AMANDA and/or IceCube contains the information about
hits in the detectors of both components.  

An important feature of
the IceCube detection system is that waveforms of the signals are captured
at 300 MHz.  
For events of high energy where many DOMs have multiple hits, use
of full waveform information can improve the reconstruction.  For example
Ref.~\cite{Grullon} finds $\delta E /E\sim 0.34$ and $\delta\psi\sim0.6^\circ$
for thoroughgoing tracks with $0.1 < E_\nu < 30$~PeV.

\begin{figure}[t]
    \includegraphics[width=0.5\textwidth]{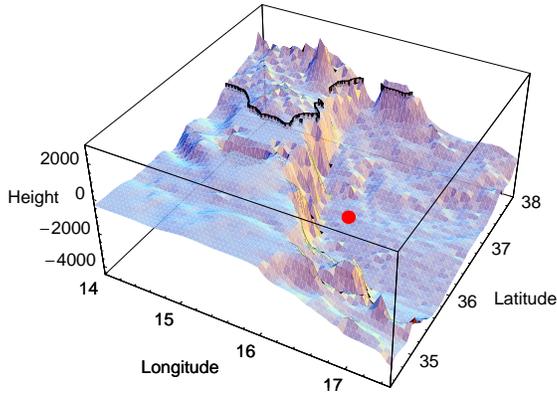}
    \caption{Surface map of the coast and sea floor of Eastern
Sicily~\cite{Cuoco}.  The line marks the shoreline, and the highest
peak is Mt. Etna.  The red dot indicates the proposed NEMO location.
}
    \label{Cuoco1}
\end{figure}

\begin{figure}[b]
    \includegraphics[width=0.5\textwidth]{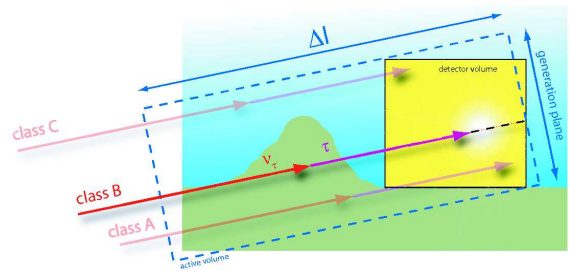}
    \caption{Diagram showing several trajectories of neutrino
interactions that pass through the sensitive volume of Auger
from below the horizon~\cite{Gora}.
}
    \label{Andes}
\end{figure}

KM3Net is a consortium of the three Mediterranean experiments (ANTARES~\cite{Kouchner}, 
NEMO~\cite{Taiuti} and NESTOR~\cite{Nestor}) to design and build a kilometer-scale
neutrino telescope in the Mediterranean.  Several papers~\cite{Carr1,White}
present studies of the configuration and sensitivity of a gigaton detector in the
Mediterranean Sea.  Of particular interest is the comparison of three
sites~\cite{Miele,Cuoco} for their sensitivities to neutrinos with
energies $E_\nu > 0.1$~EeV, the energy range of cosmogenic neutrinos.  In this
case, neutrino-induced $\tau^\pm$ and $\mu^\pm$ tracks passing through the
detector are of comparable importance.  The $\nu_\tau$ channel is enhanced
by the possibility of interaction in nearby mountains from which
the $\tau$-lepton can emerge and pass through the detector (see 
Fig.~\ref{Cuoco1}).
The geometrical and kinematical situation are complicated by
the competition among $\nu_\tau$ interaction and regeneration in
the Earth, by $\tau$ energy loss and decay and by the configuration
and response of the detector~\cite{Zas}.  Accounting for these
complications, the
estimated rate is~\cite{Miele} 0.1 GZK neutrino interactions per km$^3$ per year
in a deep sea detector.  

Because the Earth becomes opaque to neutrinos with $E_\nu\sim$PeV
the angular distribution of neutrinos from below can in principle be
used to measure the neutrino cross section by varying the path length
through the Earth, corresponding to the nadir angle of the event.
Given a sufficiently large flux of neutrinos it is conceivable to
separate neutrino cross section from energy spectrum.
Ref.~\cite{Miele} give
examples of energy/angular dependence in the energy range 10 PeV to 1 EeV.

{\bf Giant air shower arrays as neutrino detectors.}

Auger and other arrays can be used to look for horizontal air showers
initiated near the detector by a neutrino interaction with $E_\nu> 1$~EeV.
Signature of a neutrino is a horizontal shower 
observed at the ground with a large electromagnetic component
and time structure like that of a normal (nearly vertical)
cosmic-ray shower.  In contrast, the background of cosmic-ray
induced horizontal showers consists mostly of muons with a
sharp time structure of the shower front.  The electromagnetic
component of a horizontal air shower is absorbed far from the
detector because of the large slant depth.  

Horizontal showers from neutrinos are considered in two groups~\cite{Zas},
those from above the horizon, which are dominated by charged current
interactions of $\nu_e$, and those from below the horizon, which are mainly
from $\nu_\tau$ that interact in the surface of the Earth and produce
a $\tau$-lepton that emerges from the ground and decays in
the atmosphere in the field of view of the detector.  The present
limit from early operation of Auger is cited above (see Fig.~\ref{GZK}).  
A detailed Monte Carlo simulation of Auger South to GZK
neutrinos, accounting for the nearby Andes mountains, is given in Ref.~\cite{Gora}
and illustrated in Fig.~\ref{Andes}.  The contribution of Earth-skimming $\nu_\tau$
is enhanced by the presence of the Andes mountains.  In Fig.~\ref{Andes} lines 
labeled A and B would be from decay of 
$\tau$-leptons, while C most often would be from the interaction of a $\nu_e$.  
Class C events can also arrive from above the horizon.

{\bf Radio detection of neutrinos.}
Given the low event rates expected from cosmogenic neutrinos
in optical and air shower detectors, and in view of the
importance of the measurement, efforts to find techniques
that allow a much larger effective volume for detecting 
neutrinos in the EeV energy range are important.  Using radio Cherenkov radiation
from neutrino interactions in ice is one way to do this~\cite{Jaime}.
The Radio Antarctic Muon and Neutrino Detector (RAMAND) at the Vostok station
was the first effort to investigate radio detection in ice~\cite{RAMAND}.
In a paper at this conference~\cite{RadioSims} some of the originators
of the radio technique report on a hybrid Monte Carlo code (SIMEX) that allows
fast simulations of radio Cherenkov radiation from neutrino interactions~\cite{RadioSims}.
The RICE detector at the South Pole~\cite{RICE2} 
is still in operation using the same technique.
Studies of radio detection in ice at 
the South Pole continued in the 2006-2007 season
with the deployment of test receivers and transmitters by AURA~\cite{Landsman}.

Another approach is to use a balloon-borne detector looking down at
the Antarctic ice sheet.  Results from a prototype flight of ANITA
are included in Fig.~\ref{GZK}.  The full ANITA detector flew over 
Antarctica
for 35 days after launch on December 15, 2006~\cite{Palladino}.  Results have not
yet been reported.  Before the flight, the detector was placed above a large
block of ice illuminated by an intense, pulsed electron beam at SLAC~\cite{ANITAtest}.
Measurements of radio pulses confirm the theory of Askaryan~\cite{Askaryan} on which
interpretation of measurements in the field will be based.
Aspects of propagation of radio signals through the ice-snow-surface interface
were checked by observing signals broadcast from a bore hole in the Ross
Ice Shelf while ANITA was still in sight~\cite{Goldstein}.  The possibility
of using air shower cores to calibrate a radio array in the field was
discussed in Ref.~\cite{Seckel}.

\begin{figure}[t]
\includegraphics [width=0.33\textwidth, angle=270]{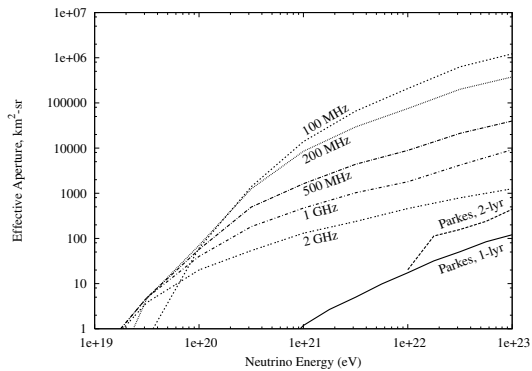}
    \caption{Aperture of the Square Kilometer Array for
detection of neutrino interactions in the Moon~\cite{Parkes}.
}
    \label{ParkesFigure}
\end{figure}

ARIANNA~\cite{ARIANNA} is a proposal to use an array of 10,000 antennas spread over
1000 km$^2$ just below the snow surface
to detect radio signals of neutrinos that interact in the ice.
Signals of downward events would be reflected by the ice-water interface.

The history and prospects for using the Moon as the target and large radio telescopes
on Earth as the detector is reviewed in Ref.~\cite{Parkes}.  Figure~\ref{ParkesFigure}
shows the calculated apertures for various frequencies for the Square Kilometer Array~\cite{SKA}.
The apertures for the original measurement with the Parkes telescope~\cite{Parkes1}
are also shown.  Techniques for pulse detection and event reconstruction
are discussed in Ref.~\cite{PulseMethods}.  Although the effective areas
achievable can be very large for SKA, as shown in Fig.~\ref{ParkesFigure},
the energy threshold is well above the range for GZK neutrinos. 

{\bf Acoustic detection of neutrinos.}  This technique is being explored by measurements
in the ice at the South Pole as another approach to achieving the effective volume
needed to measure the spectrum of cosmogenic neutrinos~\cite{SPATS}.  There is also an acoustic test setup
in Lake Baikal~\cite{BaikalAcoustic}.  Exploration of acoustic detection in ice is
motivated by the lower noise rate in ice as compared to water, which should allow
a lower energy threshold for neutrino detection.  The goal is to instrument a sufficiently
large volume to allow the detection of hundreds of GZK neutrinos per year.  The authors
suggest a hybrid approach~\cite{hybrid} using acoustic, radio and optical detectors to optimize the 
sensitivity and acceptance of the detector.

{\bf Acknowledgments} This work is supported in part by a grant
from the National Science Foundation.  I am grateful to Albrecht Karle
for helpful comments on the first draft of this paper.


\end{document}